# Nanoscale Effects on Heterojunction Electron Gases in GaN/AlGaN Core/Shell Nanowires

Bryan M. Wong[1], François Léonard*[1], Qiming Li[2], and George T. Wang[2]

[1]Sandia National Laboratories, Livermore, California, 94551

[2]Sandia National Laboratories, Albuquerque, New Mexico, 87185

*Corresponding author. E-mail: fleonar@sandia.gov.

ABSTRACT

The electronic properties of heterojunction electron gases formed in GaN/AlGaN core/shell nanowires with hexagonal and triangular cross-sections are studied theoretically. We show that at nanoscale dimensions, the non-polar hexagonal system exhibits degenerate quasi-one-dimensional electron gases at the hexagon corners, which transition to a core-centered electron gas at lower doping. In contrast, polar triangular core/shell nanowires show either a non-degenerate electron gas on the polar face or a single quasi-one-dimensional electron gas at the corner opposite the polar face, depending on the termination of the polar face. More generally, our results indicate that electron gases in closed nanoscale systems are qualitatively different from their bulk counterparts.

Bulk semiconductor-semiconductor heterojunctions have been instrumental in enabling technological breakthroughs in electronics and optoelectronics. Perhaps their biggest impact in science and technology has been through the ability to create a two-dimensional electron gas (2DEG) at these heterojunctions, which has allowed the development of high electron mobility transistors[1] and the detailed study of fundamental physics in low-dimensional correlated electron systems[2].

While further reduction of the dimensionality towards quasi-one-dimensional electron gases (Q1DEGs) has been explored through electrostatic control[3], a structurally-confined Q1DEG would have many advantages, including compactness and the promise of assembly into complex three-dimensional architectures. Recently, a path towards this goal of free-standing Q1DEGs has emerged through the synthesis of core/shell nanowires[4-10]. These systems are believed to lead to Q1DEGs in two ways: in the first case relevant to Ge/Si[5], the shell serves as a potential barrier and the electron density is confined to the core; in the second case relevant to III-V systems such as GaN/AlGaN[6,10], one may expect an electron gas to form *directly at the interface* between the core and shell, much like the case of bulk III-V heterojunctions. While this latter type of heterojunction has been extensively studied in bulk materials and is now well understood, the case of nanowire core/shell heterojunctions is much more complex because of the large parameter space that is available to control the nanowire properties. For example, the bandgaps, band offset, composition, sizes, and doping of the core and shell are all parameters that can influence the electronic properties; and while the bulk system is invariant with respect to inversion of these parameters across the interface, in a nanowire the core and shell are not geometrically equivalent. In addition, the cross-sectional geometry is a new additional parameter in the nanowire systems. Theory and modeling provide an approach to not only explore this large parameter space but also to bring a fundamental understanding of the basic electronic properties of these novel nanomaterials. Existing work in this area has focused mainly on the situation relevant to Ge/Si[11-13].

Here we present a study of the electronic properties of heterojunction electron gases in core/shell nanowires with hexagonal and triangular cross-sections, focusing on the GaN/AlGaN system, which is known to form a 2DEG in bulk heterojunctions. We find that the nanometer size combined with the highly anisotropic cross-section strongly influences the behavior of the EG, leading to confinement at corners and polar faces, and transitions between core-centered and interface-confined EGs.

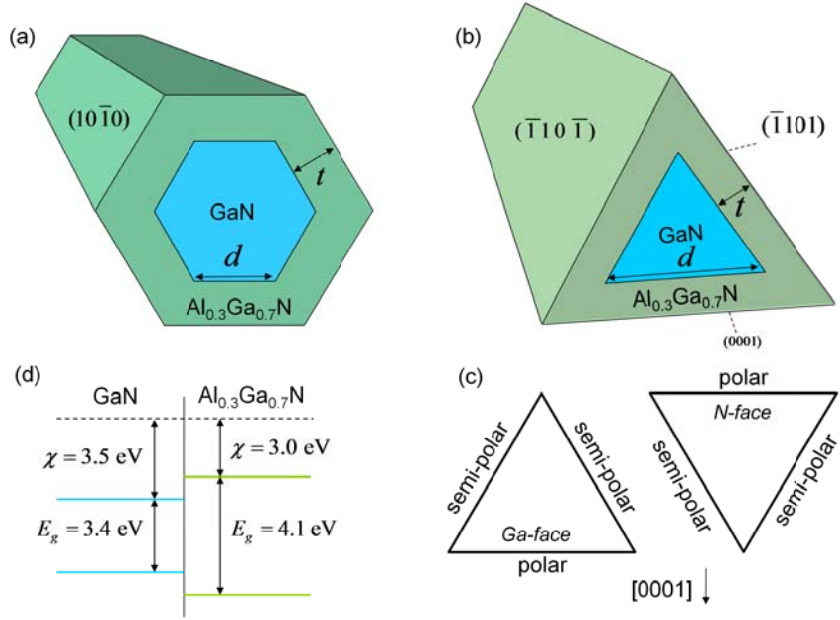

**Figure 1:** (a) and (b) show schematics of the hexagonal and triangular GaN/AlGaN core/shell nanowires considered in this work. Panel (c) shows the two variants of the triangular geometry. Panel (d) is the band alignment at the core/shell interface.

We first describe the systems considered in our calculations and illustrated in Fig. 1. The nanowires consist of a hexagonal or triangular GaN core of side length $d$ surrounded by an $Al_{0.3}Ga_{0.7}N$ shell of uniform thickness $t$ (while we focus on this particular material combination in this paper for concreteness, the main qualitative results are expected to apply to a broad range of compositions). The hexagonal and triangular geometries correspond to the ones observed experimentally[4,6,10,14,15]. The hexagonal nanowire has its axis in the [0001] direction and the cross-section is bounded by $\{10\bar{1}0\}$ planes. For the triangular nanowire, the axis direction is $[11\bar{2}0]$ and the cross-section is bounded by two equivalent $(\bar{1}101)$ and $(\bar{1}10\bar{1})$ planes and a $(0001)$ plane. It is important to mention that the triangular case has two variants depending on the orientation of the $(0001)$ plane, i.e. in the $[0001]$ or $[000\bar{1}]$ direction, as illustrated in Fig. 1c. These are often referred to as the Ga-terminated or N-terminated faces, respectively. In all cases we assume that the core/shell nanowires are defect-free, since this has been observed experimentally[6,16].

The electronic properties of the core and shell are taken to be those of the respective bulk systems. For $Al_xGa_{1-x}N$, the bandgap, affinity, dielectric constant, and isotropic effective mass satisfy the relationships[17] $E_g(x) = 3.42 \text{ eV} + x2.86 \text{ eV} - x(1-x)1.0 \text{ eV}$, $\chi = 5.88 \text{ eV} - 0.7 E_g(x)$, $\varepsilon(x) = 9.28 - 0.61x$, and $m^*(x) = (0.20 - 0.12x) m_0$. For the particular compositions x=0 and x=0.3 considered here, this leads to a Type I (straddling) heterojunction with a conduction band discontinuity $\Delta E_c = 0.5 \text{ eV}$, as shown in Fig. 1d. (We note that the quantum confinement can modify the bandgap. In planar junctions, typical bandgap corrections due to quantum confinement are on the order of 10%[18]. Given the nanowire core sizes studied here, similar small corrections are expected; these would not affect the qualitative results of the paper, namely the significant changes in the types of electron gases that are formed because of geometry.)

To obtain the electronic properties, we use a finite element, self-consistent Poisson-Schrödinger approach. In the effective mass approximation, Schrödinger's equation is

$$\left[ -\frac{\hbar^2}{2} \nabla \cdot \frac{1}{m(\vec{r})} \nabla - eV(\vec{r}) + V_{xc}(\vec{r}) \right] \psi_n(\vec{r}) = E_n \psi_n(\vec{r}) \qquad (1)$$

where $\psi_n(\vec{r})$ is the electron wave function for state $n$, $E_n$ its energy, $V(\vec{r})$ the electrostatic potential, and $V_{xc}(\vec{r})$ the electron-electron exchange-correlation potential within the local density approximation (LDA).[19] We write the wave function in the form $\psi_n(\vec{r}) = \sum_k e^{ikz} \theta_{nk}(x,y)$ where $k$ is the wavevector along the axis of the nanowire; $\theta_{nk}(x,y)$ satisfies

$$\left[-\frac{\hbar^2}{2}\vec{\nabla}\cdot\frac{1}{m(x,y)}\nabla - eV(x,y) + V_{xc}(x,y) - \frac{\hbar^2 k^2}{2m(x,y)}\right]\theta_{nk}(x,y) = E_n\theta_{nk}(x,y). \quad (2)$$

The electronic structure of the core/shell nanowire thus consists of subbands of index $n$ given by the energy dispersion $E_n(k)$. All of our calculations presented here include exchange-correlation effects through $V_{xc}(\vec{r})$, but we found the results to be similar to uncorrelated calculations without $V_{xc}(\vec{r})$, in agreement with previous studies in GaN/AlGaN heterojunctions.[20]

Schrödinger's equation is coupled to Poisson's equation through $V(\vec{r})$ and $\psi_n(\vec{r})$:

$$\nabla\cdot\left[\varepsilon(\vec{r})\nabla V(\vec{r}) + \mathbf{P}(\vec{r})\right] = -\rho(\vec{r}) = \frac{2e}{\pi}\sum_n |\psi_n(\vec{r})|^2 \int_0^\infty dk\, f(E_n - E_F) + eN(\vec{r}) \quad (3)$$

where $\mathbf{P}(\vec{r})$ is the polarization, and $f$ is the Fermi distribution with Fermi level $E_F$. The first term on the right-hand side is the charge originating from occupation of the wavefunctions obtained by solving Eq. (2). $N(\vec{r})$ is the spatially-dependent free carrier concentration due to dopants. This term affects the first because it impacts the potential $V(\vec{r})$, and thus $\psi_n(\vec{r})$ through solution of Eq. (2). We perform our calculations at zero temperature. Equations (1), (2), and (3) are augmented by appropriate boundary conditions and constraints. We set $\psi(\vec{r}) = 0$ at the surface of the nanowire. In addition, we need the position of the Fermi level. For the hexagonal case we use the charge neutrality condition $\int d\vec{r}\,\rho(\vec{r}) = 0$ to adjust the Fermi level at each iteration step. For the triangular case, we use the fact that the AlGaN/vacuum interface contains a high density of surface states that counterbalance the large spontaneous polarization charge generated at the interface[21]. The net effect is to pin the Fermi level in the AlGaN bandgap. In thin films of AlGaN on GaN, the position of the Fermi level depends on the AlGaN thickness, locating it between 1 eV and 1.8 eV below the conduction band edge[21]. For the 20-nm AlGaN shell thicknesses considered in the present paper, the value is around 1.65 eV, which we use in all of our calculations to set a boundary condition on the electrostatic potential at the surface.

To numerically implement the self-consistent calculation for arbitrary cross-section geometries, we use a finite element approach for both the Schrödinger and Poisson equations. First, we discretize the nanowire cross-section using a Delaunay triangulation method which creates a flexible grid of evaluation points. For each geometry, we use a dense grid consisting of 40,000 triangular elements which we found necessary to accurately describe the highly-localized and oscillatory wave functions in our polar interfaces. The solutions of both the Schrödinger and Poisson equations are expanded in the basis of the triangular mesh points, yielding either a sparse eigenvalue equation or a large system of linear equations, respectively. Schrödinger's equation is solved using an iterative Arnoldi algorithm to simultaneously obtain the eigenvalues and eigenfunctions, and the solution to Poisson's equation is determined by inverting the large sparse coefficient matrix. The resulting solution of our first iteration yields a potential energy which is then re-inserted into a new Schrödinger equation. The iteration procedure is repeated until self-consistency is achieved, which we choose to be a 0.01 eV average energy difference across all nodes of the electrostatic potential between successive iterations. Using our dense triangular grid, self-consistency between the Schrödinger and Poisson equations is typically achieved with less than 80 iterations.

In the GaN/AlGaN system, the polarization $\mathbf{P}(\vec{r})$ comes from two sources: the spontaneous polarization due to the polar nature of interfaces, and the piezoelectric polarization due to the strain created by the lattice mismatch. Both of these lead to a net charge density at the interface due to the discontinuity in $\mathbf{P}(\vec{r})$. For GaN and AlGaN, the spontaneous polarization is given by $\mathbf{P}^{sp} = P^{sp}\hat{z}$ where $\hat{z}$ is a unit vector in the [0001] direction. The charge at a GaN/AlGaN interface is thus $\sigma = \nabla \cdot \mathbf{P} = x\left(P^{sp}_{GaN} - P^{sp}_{AlN}\right)\cos\theta$ where $\theta$ is the angle of the interface with respect to the [0001] direction. This angular dependence implies that the interfaces in the hexagonal cross-section are all non-polar, while the triangular geometry has one polar and two semi-polar faces. We assume that the spontaneous polarization satisfies $P^{sp}_{Al_xGa_{1-x}N}(\vec{r}) = (1-x)P^{sp}_{GaN} + xP^{sp}_{AlN}$ and the values for GaN and AlN are taken from Ref. 22: $P_{GaN}^{sp} = -0.029\ cm^{-2}$ and $P_{GaN}^{sp} = -0.081\ cm^{-2}$. For the finite-element calculations, we distribute the interface charge as a Gaussian around the interface with a width at half-maximum of 2 nm.

We calculate the piezoelectric polarization at the interfaces $\mathbf{P}^{pz}$ from the relation $\mathbf{P}^{pz} = \left[ e_{15}\varepsilon_{xz}, e_{15}\varepsilon_{yz}, e_{31}(\varepsilon_{xx}+\varepsilon_{yy}) + e_{33}\varepsilon_{zz} \right]$ where $e_{ij}$ is the piezoelectric tensor, with components obtained from Ref.17. The strains in the core and shell come from the lattice mismatch between GaN and AlGaN; one can calculate the three-dimensional strains in the core and shell using, for example, equilibrium continuum elasticity, as was recently done for a cylindrical core/shell NW geometry for Si/Ge[23] or for the hexagonal core/shell NW GaN/AlN system[10]. However, these calculations show that the strain discontinuity at the interface is the same as that of a planar film, and that the strain gradients in the shell are much less than those right at the interface. Thus, we neglect the volume piezoelectric polarization in the shell, and concentrate on the interfacial polarization. This is obtained from the expressions for thin films[22] for each interface orientation. Furthermore, because the strain is relatively low in the structures considered here (less than 1%), we neglect the impact of strain on bandgaps and effective masses since those would only be changed by a few percent according to calculations on bulk GaN[24].

**Hexagonal cross-section.** We first discuss the results of our calculations for non-polar core/shell nanowires of hexagonal cross-section. In the hexagonal system of Fig. 1, the spontaneous polarization charge vanishes because the polarization axis is in the axial direction. The piezoelectric polarization charge vanishes as well because the strain components $\varepsilon_{xz}$ and $\varepsilon_{yz}$ are both zero since the displacements are uniform in the axial direction. Thus, the formation of an electron gas at the core/shell interface is entirely due to the band alignment.

We first consider core and shell n-type doping equal to $2 \times 10^{17}\ cm^{-3}$ (the results also apply to p-type doping but with hole accumulation instead). Figure 2 shows the calculated electron density and band-bending for a nanowire of core side length $d = 20$ nm and shell thickness $t = 20$ nm. A priori, one might have expected to observe an electron gas of uniform density along the faces of the nanowire, in analogy with the bulk situation. However, the nanowire case behaves much differently showing instead that six degenerate Q1DEGs are formed at the corners.

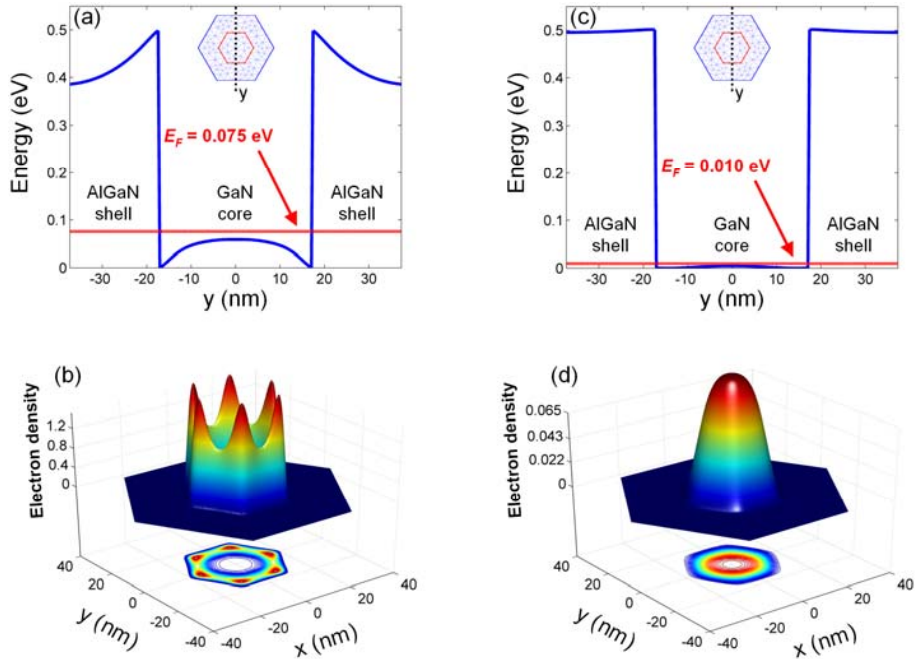

**Figure 2:** Calculated band-bending (a) and charge distribution (b) along the dashed line in the inset for a hexagonal core-shell nanowire of 20 nm core size and doping $2\times10^{17}$ cm$^{-3}$. Panels (c) and (d) show the same for a doping of $1.2\times10^{16}$ cm$^{-3}$.

The origin of this behavior lies in the electron wavefunction for the lowest energy state. Indeed, as shown in Fig. 3, the lowest energy mode consists of a high electron density at the six corners. Higher energy modes give rise to some weight along the interface, leading to a non-zero electron distribution on the faces of the core/shell nanowire.

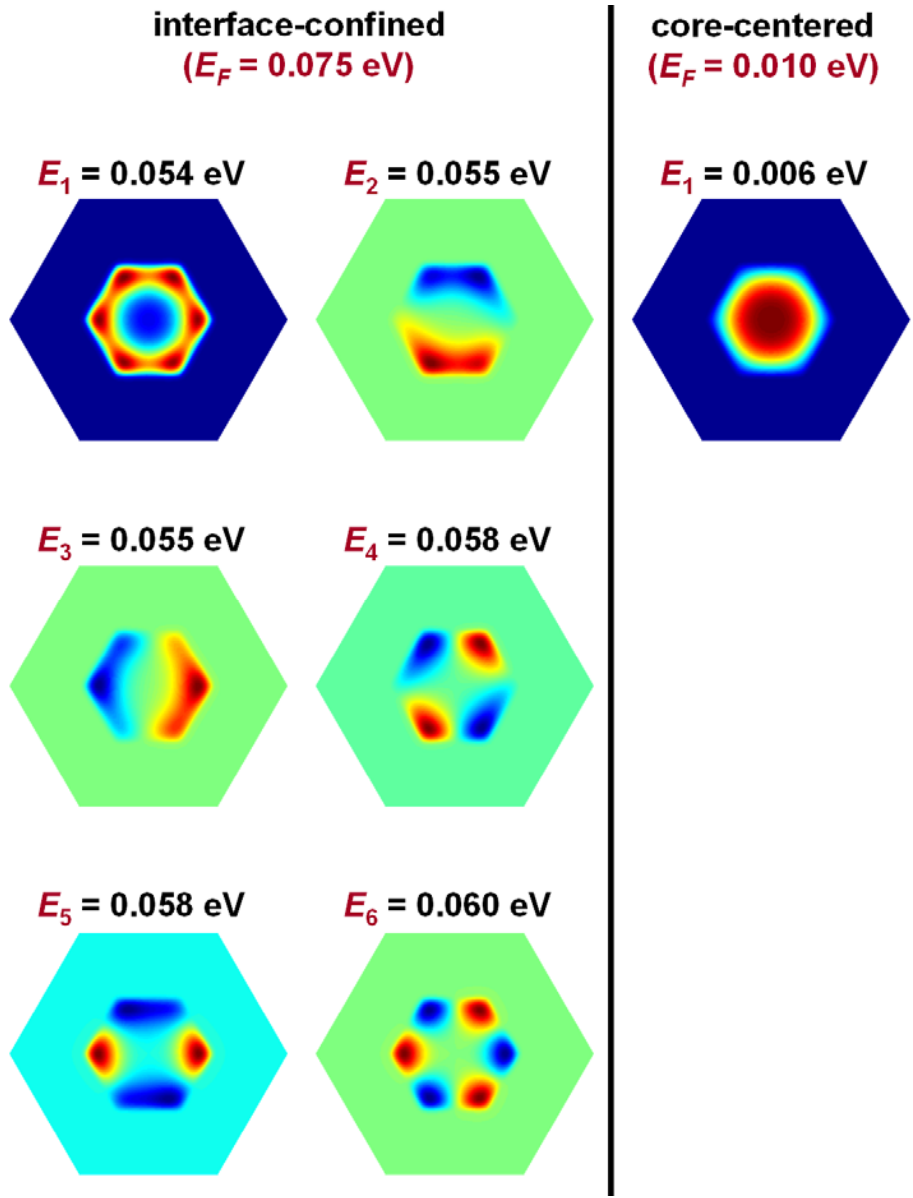

**Figure 3:** Electron wavefunctions for the lowest energy modes in the hexagonal core/shell nanowire for the interface-centered and core-centered cases. Energies measured from bottom of conduction band.

At lower doping, the situation changes qualitatively. Because screening lengths increase with decreasing doping, at low doping the band-bending at the heterojunction on the core side becomes comparable to the core size, leading to a flat potential inside the core. Thus, confinement occurs in the whole core instead of at the interface, as shown in Fig. 2c. The electron density of the lower energy state peaks at the core center, giving a qualitatively different type of EG (Fig. 2d). To quantify the transition between the interface-confined and core-centered EG, we calculated the average location of the wavefunction maximum for the lowest energy state as a function of both core size and doping. We define the core-centered (interface-centered) regime when the maximum is less (greater) than 10% (80%) of the core size away from the center of the nanowire. Figure 4 shows the transition between the two regimes as a function of core size and doping. The figure indicates that the interface-centered EG requires relatively high doping, especially as the core size is reduced. Furthermore, we also find that at low doping, the lowest energy level is unoccupied, and no electron gas exists.

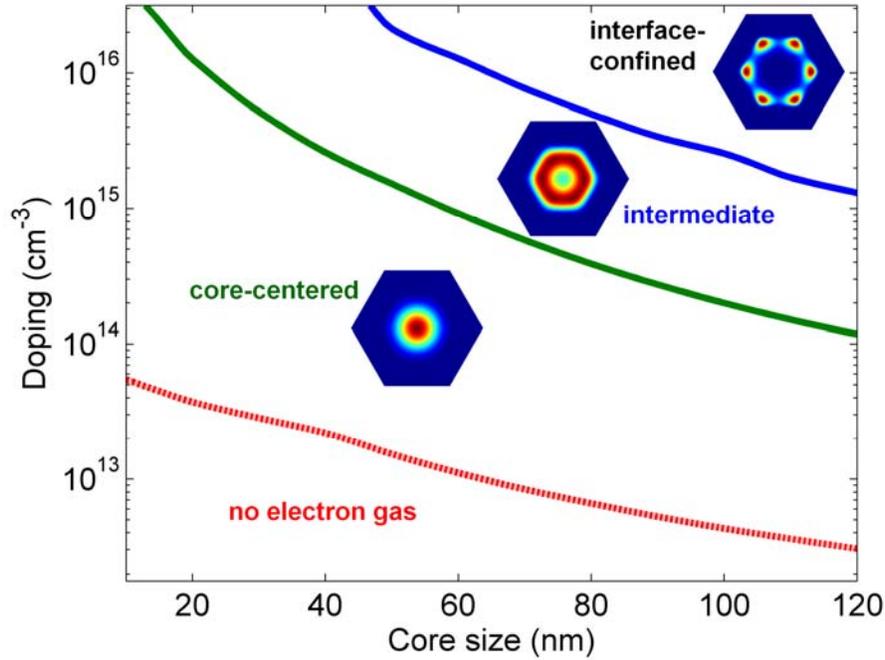

**Figure 4:** Diagram showing the three types of electron gases found in the hexagonal core-shell nanowires as a function of core size and doping.

**Triangular cross-section.** The triangular geometry presents an interesting situation because of the presence of both spontaneous and piezoelectric polarization, and because of the two possible variants of Fig. 1c. We first discuss the case of the Ga-face orientation with n-type doping in both the core and shell (the results also apply to p-type doping but with hole accumulation instead), where the spontaneous polarization creates a positive charge at the (0001) interface, and a negative charge at the two semi-polar faces. The free electrons due to the n-type doping are attracted to the (0001) interface, thus creating a 2DEG there. Figure 5 shows the charge distribution for four different core sizes. The distribution evolves from a highly-peaked structure to a sheet as the core size increases. This originates from the quantized states and their wavefunctions at the interface, which show peaks and nodes; the number of occupied states determines the shape of the charge distribution. As the core size increases, the energy difference between the quantized states decreases, and the charge distribution approaches that of a thin film.

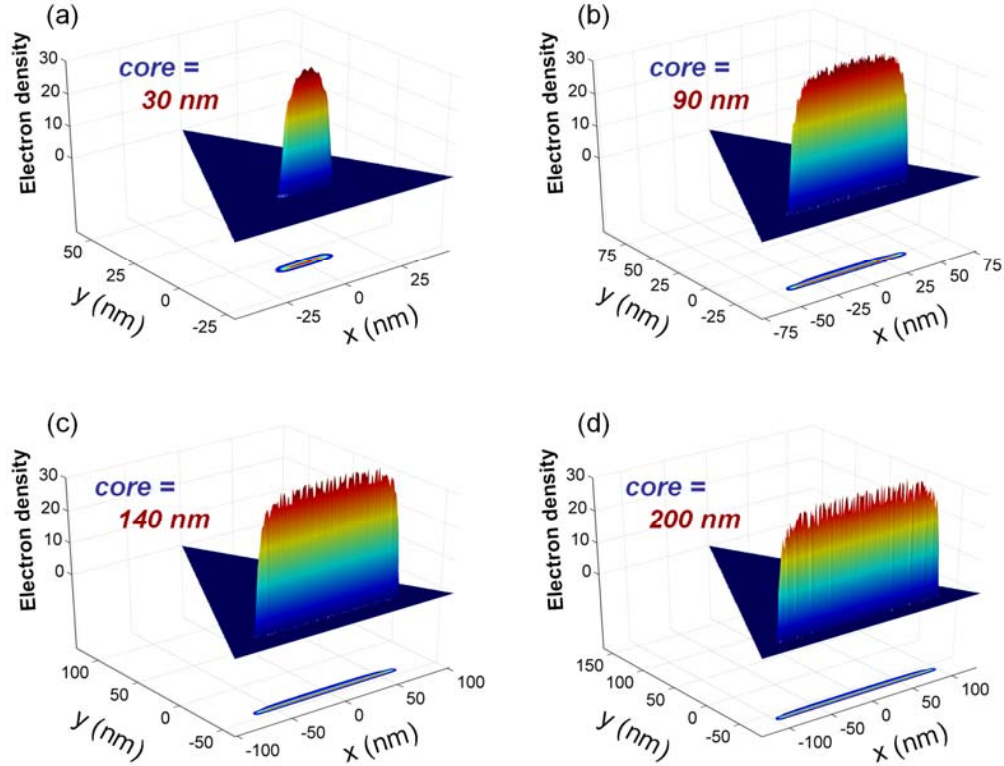

**Figure 5:** Charge distribution for the (0001) Ga-face triangular core/shell nanowire for four core sizes.

The case of the N-face system shows a qualitatively different behavior. Because the $(000\bar{1})$ interface has negative polarization charge, the free electrons are repelled from that interface, but attracted to the other two faces due to the positive polarization charges there. The system reaches a compromise by creating an electron gas that is localized near the corner of the triangle, as shown in Fig. 6. As the core size increases, the electron gas extends along the two semi-polar faces since the corner is farther away from the negatively polarized $(000\bar{1})$ interface. Thus, the small core size system may be the one closest to forming a Q1DEG.

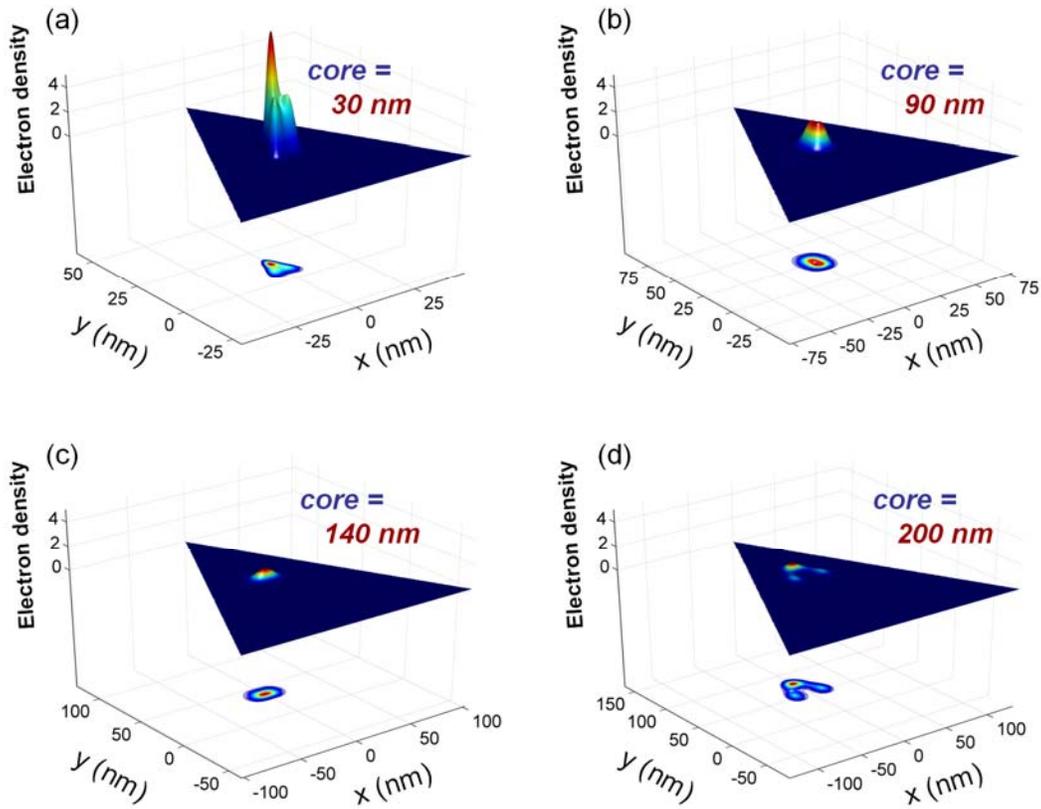

**Figure 6:** Charge distribution in the triangular $(000\bar{1})$ N-face core/shell nanowire for four core sizes.

The presence or absence of an EG in the core/shell nanowire systems depend critically on the doping and nanowire dimensions. Indeed, Fig. 7 shows that below a critical line determined by doping and core size, the Fermi level is below the lowest confined energy state, and no electron gas is present. Thus, the experimental observation of EGs in these systems requires a detailed control over geometry, dimensions, crystal orientation, and doping. For example, if an EG in nanowires with core size greater than 15 nm is desirable, the triangular Ga-face system may be one of choice since an EG exists at any doping. For smaller core sizes, EGs may be more easily realizable in the hexagonal system since relatively low doping is required to establish an EG.

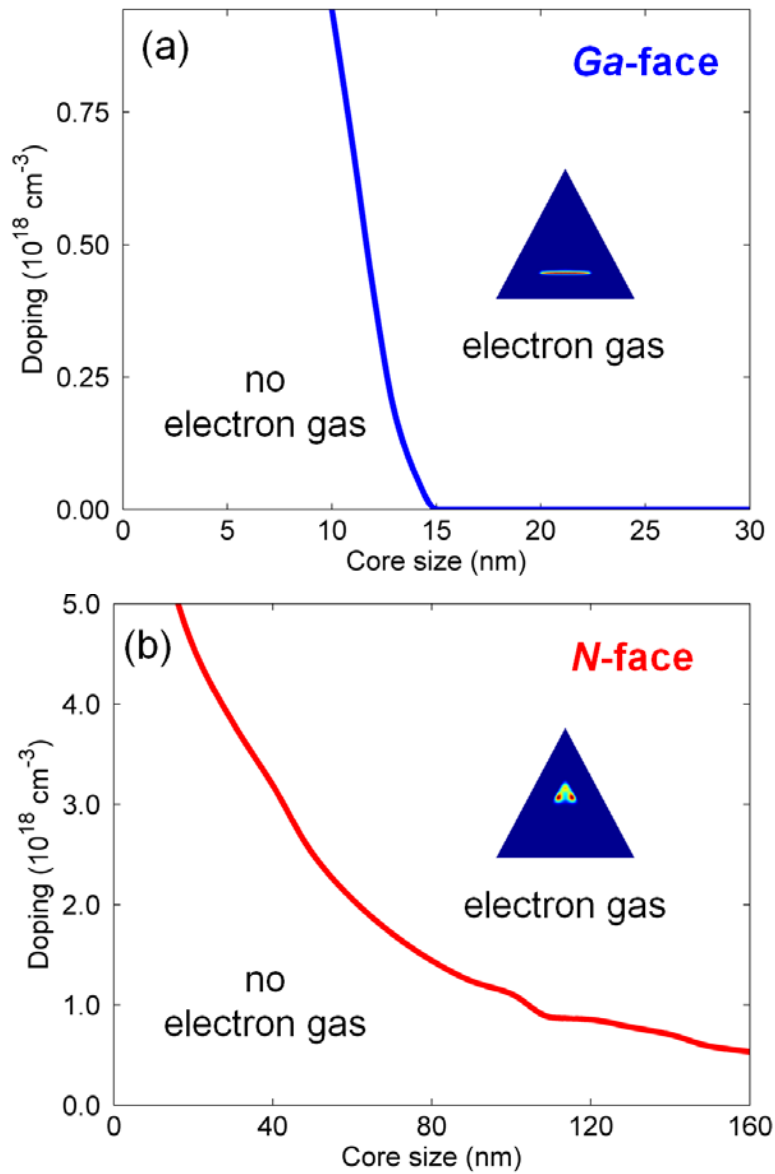

**Figure 7:** Critical doping as a function of core size below which no electron gas exists in the triangular core/shell nanowires, for the (a) Ga-face and (b) N-face systems.

In conclusion, we find that electron gases at core/shell nanowire interfaces show unusual properties compared to their bulk counterparts. The nanoscale geometry introduces new complexities that lead to novel electron localization effects. Our calculations have many implications for experiments. First, the electronic energy levels and their symmetries are specific to the different types of electron gases, which would impact electronic transport and optical experiments. Second, spatially-resolved electronic and optical experiments (e.g. catholuminescence, photoluminescence, electrical nanoprobing) could be utilized to probe the heterojunction properties. Third, the corner or face localization in the triangular case implies that careful consideration has to be given to device geometries and orientations that exploit these electron gases, including the formation of contacts. Finally, the central result that nanoscale geometry changes qualitatively carrier distributions should also impact other types of electronic and photonic devices based on core/shell nanowires, such as light-emitting diodes and solar cells.

ACKNOWLEDGMENT

This project was supported by the Solid-State Lighting Science Center, an Energy Frontier Research Center (EFRC) funded by the U.S. Department of Energy, Office of Science, Office of Basic Energy Sciences (BES), and the BES Division of Materials Science and Engineering. Additional support provided by the Laboratory Directed Research and Development program at Sandia National Laboratories, a multiprogram laboratory operated by Sandia Corporation, a Lockheed Martin Company, for the United States Department of Energy under contract DE-AC04-94-AL85000.